\documentclass[a4paper,fleqn,usenatbib]{mnras}
\usepackage[T1]{fontenc}
\usepackage{ae,aecompl}
\usepackage{amsmath}
\usepackage{amssymb}
\usepackage{graphicx}
\usepackage{mathrsfs}
\usepackage{algorithm}
\usepackage{algpseudocode}
\usepackage{color}
\usepackage{epstopdf}
\setcitestyle{notesep={ }} 
\usepackage{hyperref}
\usepackage[all]{hypcap}

\title[IFS-RedEx]{IFS-RedEx, a redshift extraction software for integral-field spectrographs: Application to MUSE data}
\author[Rexroth et al.]{Markus Rexroth$^{1}$, Jean-Paul Kneib$^{1,2}$, R\'{e}my Joseph$^{1}$, Johan Richard$^{3}$, Romaric Her$^{1,4}$\\
\\
$^{1}$ Institute of Physics, Laboratory of Astrophysics, Ecole Polytechnique F\'{e}d\'{e}rale de Lausanne (EPFL), Observatoire de Sauverny,\\ \hphantom{$^{1}$} 1290 Versoix, Switzerland\\
$^{2}$ Aix Marseille Universit\'{e}, CNRS, LAM (Laboratoire d'Astrophysique de Marseille) UMR 7326, 13388, Marseille, France\\
$^{3}$ Univ Lyon, Univ Lyon 1, ENS de Lyon, CNRS, Centre de Recherche Astrophysique de Lyon UMR5574, F-69230, Saint-Genis-Laval, France\\
$^{4}$ ISAE-SUPAERO, Universit\'{e} de Toulouse, 10 Avenue Edouard Belin, 31400 Toulouse, France
}

\date{Accepted XXX. Received YYY; in original form ZZZ}
\pubyear{2017}

\begin{document}
\label{firstpage}
\pagerange{\pageref{firstpage}--\pageref{lastpage}}
\maketitle

\begin{abstract}
We present IFS-RedEx, a spectrum and redshift extraction pipeline for integral-field spectrographs. A key feature of the tool is a wavelet-based spectrum cleaner. It identifies reliable spectral features, reconstructs their shapes, and suppresses the spectrum noise. This gives the technique an advantage over conventional methods like Gaussian filtering, which only smears out the signal. As a result, the wavelet-based cleaning allows the quick identification of true spectral features. We test the cleaning technique with degraded MUSE spectra and find that it can detect spectrum peaks down to $\text{S/N} \approx 8$ while reporting no fake detections. We apply IFS-RedEx to MUSE data of the strong lensing cluster MACSJ1931.8-2635 and extract 54 spectroscopic redshifts. We identify 29 cluster members and 22 background galaxies with $z \geq 0.4$. IFS-RedEx is open source and publicly available.
\end{abstract}


\section{Introduction}
Astrophysical research has benefited greatly from publicly available open source software and programs like SExtractor \citep{Bertin1996} and Astropy \citep{AstropyCollaboration2013} have become standard tools for many astronomers. Their public availability allows researchers to focus on the science and to reduce the programming overhead, while the open source nature facilitates the code's further development and adaptation. In this spirit, we developed the Integral-Field Spectrograph Redshift Extractor (IFS-RedEx), an open source software for the efficient extraction of spectra and redshifts from integral-field spectrographs\footnote{The software can be downloaded at \url{http://lastro.epfl.ch/software}}. The software can also be used as a complement to other tools such as the Multi Unit Spectroscopic Explorer (MUSE) Python Data Analysis Framework (mpdaf)\footnote{Available at \url{https://git-cral.univ-lyon1.fr/MUSE/mpdaf}}.\\
\\
Our redshift extraction tool includes a key feature, a wavelet-based spectrum cleaning tool which removes spurious peaks and reconstructs a cleaned spectrum. Wavelet transformations are well suited for astrophysical image and data processing \citep[see e.g.][for an overview]{Starck2006} and have been successfully applied to a variety of astronomical research projects. To name only a few recent examples, wavelets have been used for source deblending \citep{Joseph2016}, gravitational lens modeling \citep{Lanusse2016} and the removal of contaminants to facilitate the detection of high redshift objects \citep{Livermore2016}.\\
\\
The paper is designed as follows: Sections 2 and 3 present the spectrum and redshift extraction routines of IFS-RedEx. In section 4, we describe and test the wavelet-based spectrum cleaning tool. In section 5, we illustrate the use of our software by applying it to MUSE data of the strong lensing cluster MACSJ1931.8-2635 (henceforth called MACSJ1931). We summarize our results in section 6.

\section{Spectrum extraction \& catalog cleaning}
It is advantageous to combine Integral-Field Unit (IFU) data cubes with high resolution imaging, as this allows us to detect small, faint sources which might remain undetected if we used the image obtained by collapsing the data cube along the wavelength axis (henceforth called white-light image) for source detection. For example, \citet{Bacon2015} used this combination in their analysis of MUSE observations of the \textit{Hubble} Deep Field South. Therefore we exploit this case in the following, but in principle the software can be used without high resolution data. IFS-RedEx uses the center positions of stars provided by the user to align the IFU and high resolution images. It utilizes a SExtractor \citep{Bertin1996} catalog of the high-resolution data to extract the spectra and the associated standard deviation noise estimate for each source from the data cube. It extracts the signal in an area with a radius of 3 to 5 data cube pixels, depending on the SExtractor full width at half maximum (FWHM) estimate. Sources with FWHM $< 2$ high resolution pixels are discarded as these are typically spurious detections, e.g. due to cosmic rays.\\
\\
IFS-RedEx shows the user each source and extraction radius overplotted on the high resolution image and the IFU data cube. The user can now quickly examine each detection and decide to either keep it in the database or to remove it, for example because it is too close to the data cube boundary and suffers from edge effects.\\
\\
The tool also supports line emission and continuum emission catalogs. These are for example created by the MUSELET\footnote{MUSELET is part of the mpdaf package. A tutorial and the documentation are available at \url{http://mpdaf.readthedocs.io/en/latest/muselet.html}} software, which uses narrow-band images to perform a blind search for the respective signal. IFS-RedEx displays the detected sources and their extraction radius of 3 pixels on the IFU data cube. The user labels sources which cannot be used, e.g. because the signal is only a spurious detection in one pixel or it is too close to the image boundary. The spectra and noise of the good sources are automatically extracted. \\
\\
Finally, the cleaned SExtractor, line emission, and continuum emission catalogs are merged into a master catalog. In this step, the sources are displayed on the high-resolution image so that the user can decide if the MUSELET and SExtractor detections are part of the same source. This visual inspection is more reliable than an automatic association and the number of sources is typically small enough for a manual inspection in reasonable time.

\section{Redshift extraction}
\begin{figure*}
\begin{center}
\includegraphics[width=\textwidth]{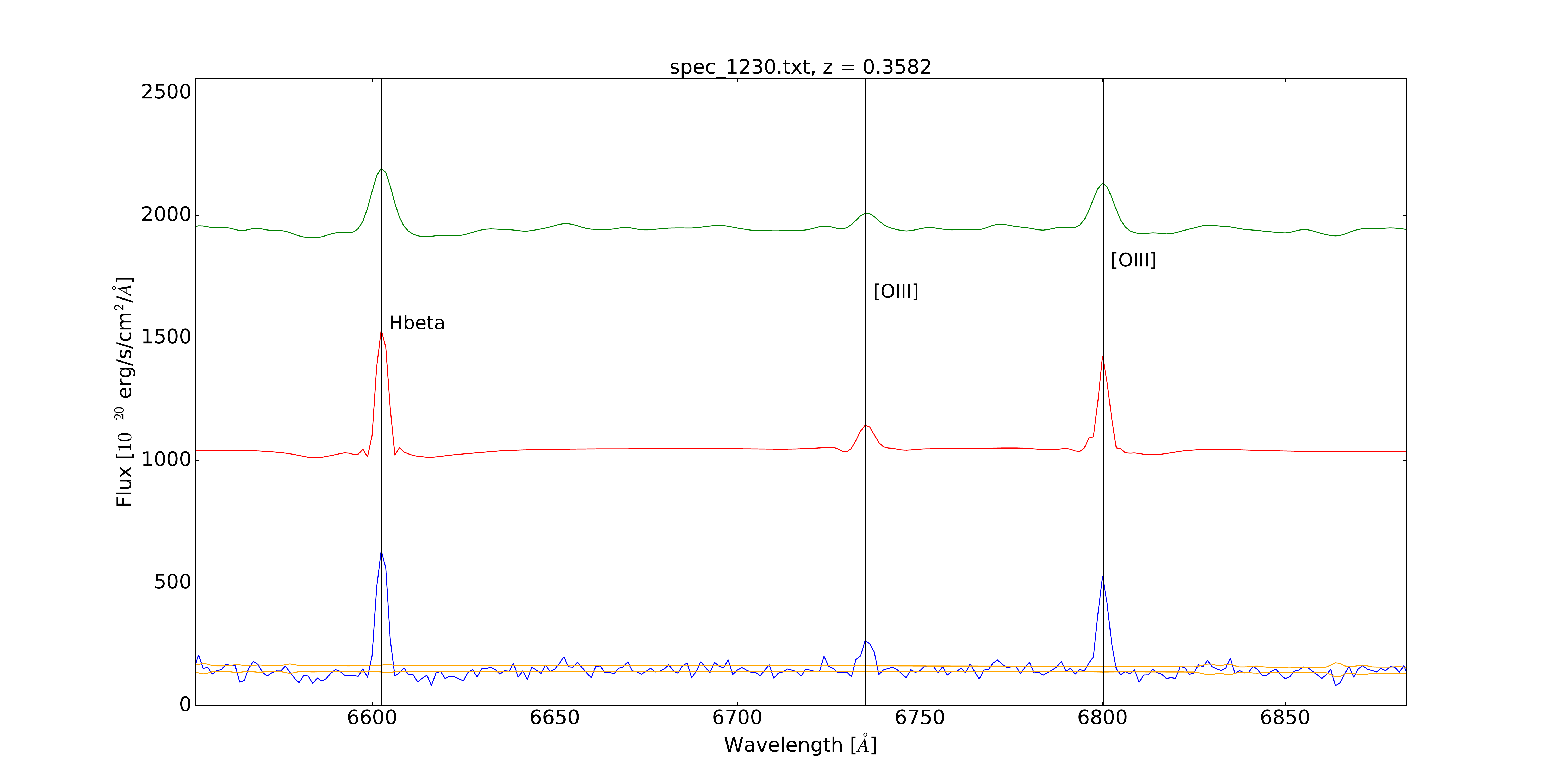}\\
\includegraphics[width=0.4838\textwidth]{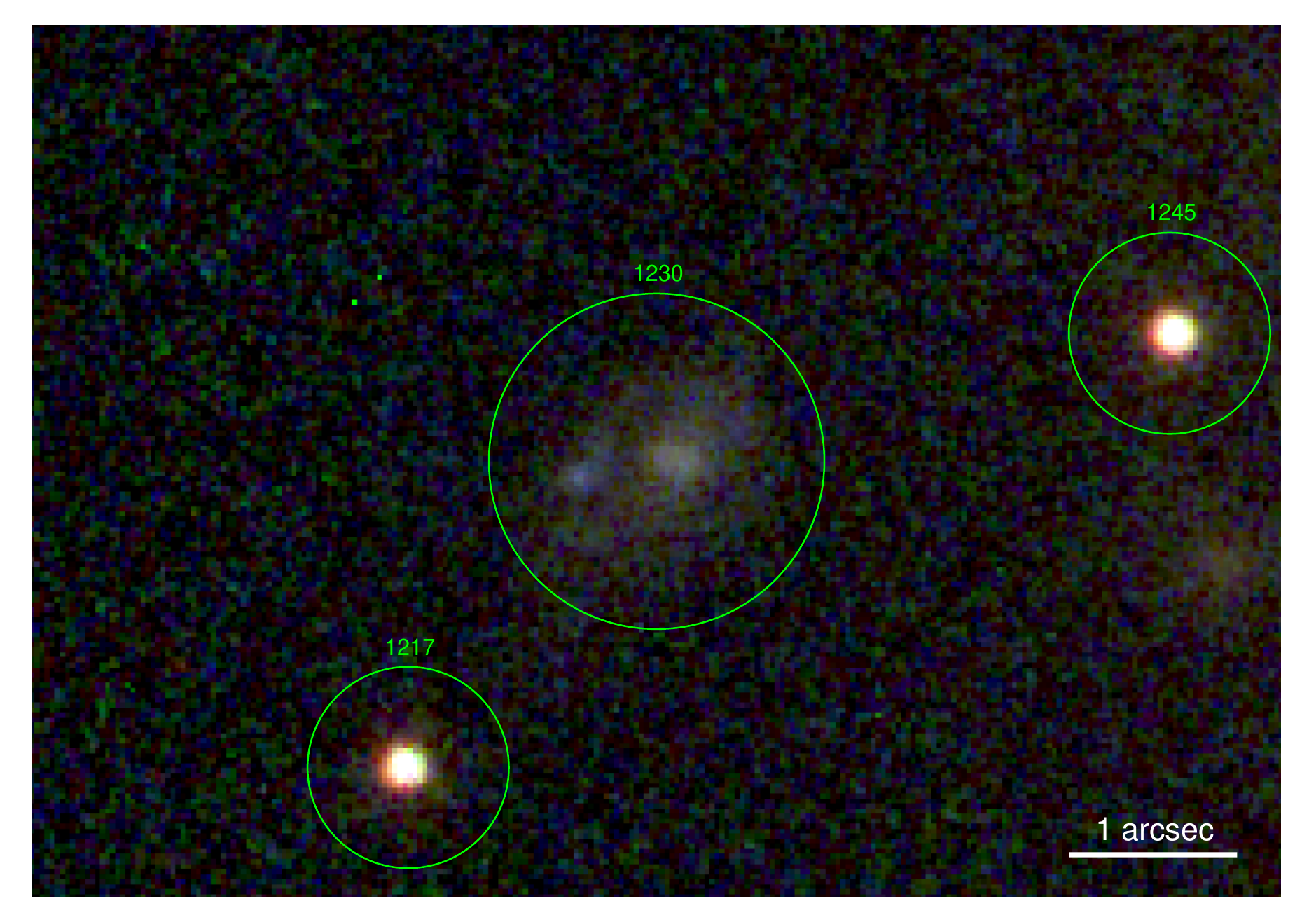}
\includegraphics[width=0.5\textwidth]{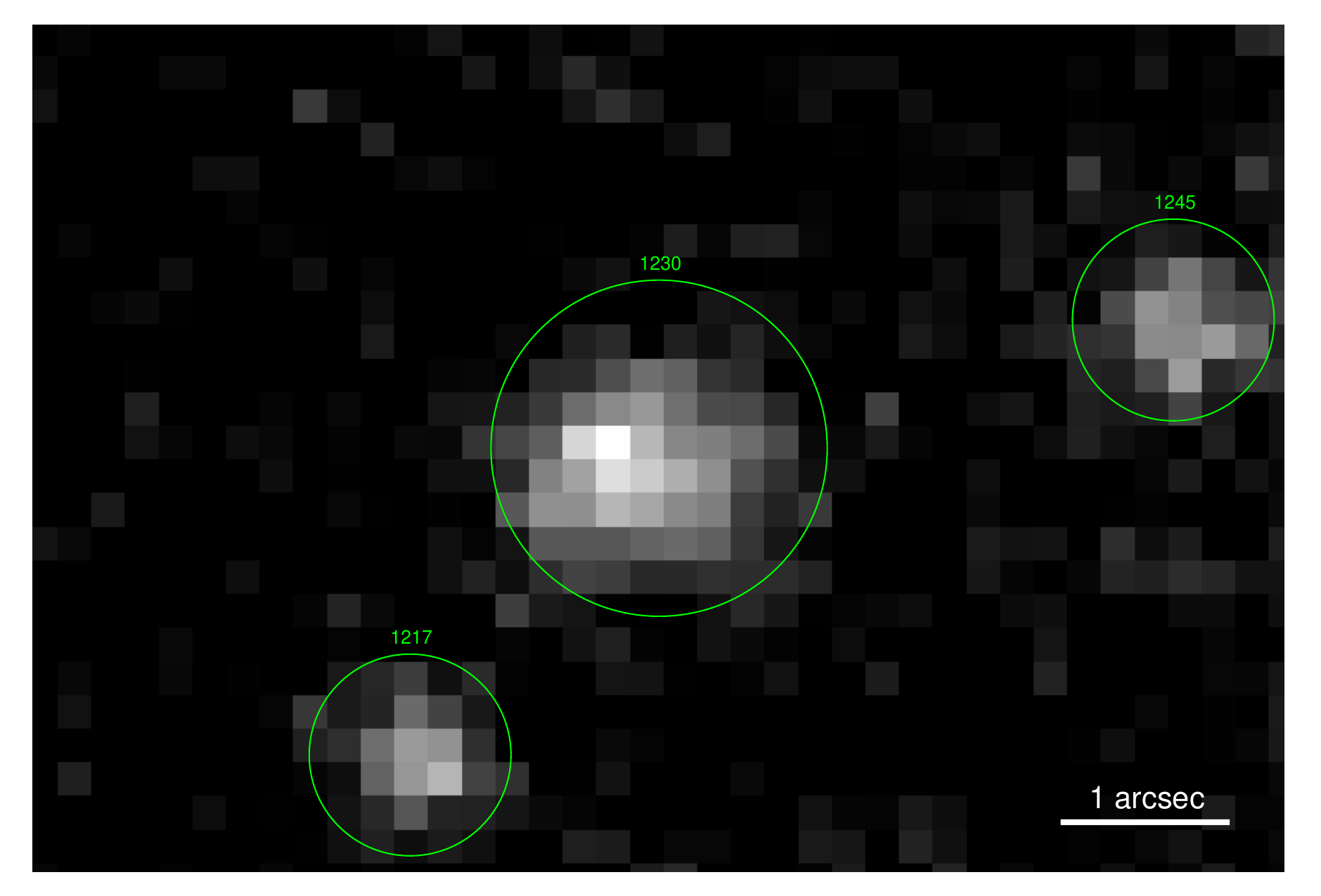}
\end{center}
\caption{Top: Interactive spectrum plot. The user can apply Gaussian filtering (green), wavelet cleaning (red) and plot the noise (yellow) to distinguish real from spurious features of the data (blue). The offsets of the plots can be adjusted. The noise shows the upper and lower standard deviation around the smoothed signal (see text).
Bottom left: IFS-RedEx displays the high resolution color image for each source to facilitate the redshift extraction. The respective source is always at the image center and labeled with the number of the source's spectrum file, here 1230. Bottom right: MUSE data cube slice at 6799.97~Angstrom corresponding to the high resolution image on the left. In the spectrum fitting step the data cube is typically not needed and thus it is not displayed by default, but it can be quickly loaded via the DS9 interface if required.}\label{fig_spectrumPlot}
\end{figure*}

Each 1D spectrum is displayed in an interactive plot and a second window shows the corresponding high resolution image, see figure~\ref{fig_spectrumPlot}. The position of sky lines with a flux $\geq 50 \times 10^{-20}~\text{erg}\, \text{s}^{-1}\,\text{cm}^{-2}\,\text{arcsec}^{-2}$ are labeled in green. The sky line fluxes are taken from \citet{Cosby2006}. IFS-RedEx also lists the emission line identifications from MUSELET if available.\\
\\
The user can now adjust the position of the emission and absorption line template by changing the source redshift. Once the template matches the source spectrum, the right redshift is found. IFS-RedEx has several features to facilitate the correct identification of spectral features. The user can zoom in and out, overplot the noise on the spectrum, smooth the signal with a Gaussian filter and perform a wavelet-based spectrum cleaning, see figure~\ref{fig_spectrumPlot}. When IFS-RedEx plots the noise, it shows the standard deviation around an offset. The offset is calculated by smoothing the spectrum signal with a Gaussian with $\sigma = 100~\text{pixels}$. Thus the noise is centered on the smoothed signal and it follows signal drifts. The wavelet cleaning is described in detail in the next section. As can be seen in figure~\ref{fig_spectrumPlot}, it reconstructs the shape of the reliable spectrum features and suppresses the noise. The Gaussian filter only smears out the signal. Thus the wavelet-based reconstruction makes it easier to distinguish true from false peaks. \\
\\
Finally, the user can fit a Gaussian to the most prominent spectral line. IFS-RedEx combines the error of the fitted center position with the wavelength calibration error from the IFU data reduction pipeline into the final statistical redshift error. The software creates a final catalog with all source redshifts and errors. In addition, it produces a document with all spectral feature identifications and high resolution images for later use, e.g. for verification by a colleague.

\section{Wavelet-based spectrum reconstruction}
\subsection{Wavelet transform algorithms}
The wavelet-based cleaning algorithm reconstructs only spectral features above a given significance threshold. For this purpose, we use the ``\`{a} trous'' wavelet transform with a $B_3$-spline scaling function of the coordinate $x \in \mathbb{R}$, 
\begin{equation}
\phi(x) = \frac{1}{12}(|x-2|^3 -4|x-1|^3 + 6|x|^3 - 4|x+1|^3 + |x+2|^3),
\end{equation}
which is well suited for isotropic signals such as emission lines \citep{Starck2007,Starck2006,Holschneider1989}. In contrast to a Fourier transform, wavelets possess both frequency and location information. We note that the measured spectrum signal is discrete and not continuous and we denote the unprocessed, noisy spectrum data $\boldsymbol{c_{0}}$, where the subscript indicates the scale $s$, and its value at pixel position $l$ with $c_{0,l}$. We assume that $c_{0,l}$ is the scalar product of the continuous spectrum function $f(x)$ and $\phi(x)$ at pixel $l$.  Now we can filter this data, where each filtering step increases $s$ by one and leads to $\boldsymbol{c_{s+1}}$, which no longer includes the highest frequency information from $\boldsymbol{c_s}$. The filtered data for each scale is calculated by using a convolution. The coefficients of the convolution mask $\boldsymbol{h}$ derive from the scaling function,
\begin{equation}
\frac{1}{2}\phi\Big(\frac{x}{2}\Big) = \sum_l h(l)\phi(x-l),
\end{equation}
and they are (1/16, 1/4, 3/8, 1/4, 1/16) \citep{Starck2006}. By noting that $h(k)$ is symmetric \citep{Starck2007}, we have
\begin{equation}
c_{s,l} = \sum_k h(k) c_{s-1,l+2^{s-1}k} \label{equation:c_sl}
\end{equation} 
and we define the double-convolved data on the same scale by
\begin{equation}
cd_{s,l} = \sum_k h(k) c_{s,l+2^{s-1}k}. \label{equation:c_sl2}
\end{equation}
The wavelet coefficients are now given by
\begin{equation}
w_{s,l} = c_{s-1,l} - cd_{s,l}, \label{equation:w_sl}
\end{equation}
and they include the information between these two scales \citep{Starck2016}. A low scale $s$ implies high frequencies and vice versa. The final wavelet transform is the set $\{\boldsymbol{w_1}, \dots, \boldsymbol{w_L}, \boldsymbol{c_L}\}$, where $L$ is the highest scale level we use, and it includes the full spectrum information. We impose an upper limit for $L$ depending on the spectrum wavelength range and resolution: $L \leq \log_2((P-1)/(H-1))$, where $P$ is the number of pixels of the spectrum signal and $H$ the length of $\boldsymbol{h}$, which is in our case $H=5$. Otherwise $s$ could become so large that the filtering equation~\ref{equation:c_sl} would require data outside of the wavelength range. We compute the wavelet transform according to algorithm~\ref{algorithm:SignalToWaveletSpace} and we transform back into real space by using algorithm~\ref{algorithm:WaveletToSignalSpace} \citep{Starck2016}.
\\
\begin{algorithm}
\renewcommand{\hypcapspace}{30pt} 
\capstart 
\renewcommand{\hypcapspace}{0.5\baselineskip} 
\begin{algorithmic}[1]
\Statex \textbf{Require:} Spectrum $\boldsymbol{c_0}$ (= set of discrete spectrum pixels $\{c_{0,l}\}$), highest scale level $L$,
convolution mask $\boldsymbol{h}$
\Statex \textbf{Output:} Wavelet transform of spectrum $\{\boldsymbol{w_1}, \dots, \boldsymbol{w_L}, \boldsymbol{c_L}\}$
\State \textbf{Procedure} \textsc{wavelet\_transform}($\boldsymbol{c_0}$, $L$)\textbf{:}
\State $s \gets 0$
\While{$s<L$}
\State $s \gets s + 1$
\State $c_{s,l} \gets \sum_k h(k) c_{s-1,l+2^{s-1}k}$~$\forall\, l$
\State $cd_{s,l} \gets \sum_k h(k) c_{s,l+2^{s-1}k} ~\forall\, l$
\State $w_{s,l} \gets c_{s-1,l} - cd_{s,l} ~\forall\, l$
\EndWhile
\State \Return $\{\boldsymbol{w_1}, \dots, \boldsymbol{w_L}, \boldsymbol{c_L}\}$
\end{algorithmic}
\caption{Transform the spectrum into wavelet space}\label{algorithm:SignalToWaveletSpace}
\end{algorithm}
\\
The cleaning in wavelet space is performed following \citet{Starck2006}: We transform a discretized Dirac $\delta$-distribution to obtain the wavelet set $\{\boldsymbol{w_1^{\delta}}, \dots, \boldsymbol{w_L^{\delta}}\}$. Subsequently, we convolve each squared $\boldsymbol{w_s^{\delta}}$ with the squared standard deviation spectrum noise extracted from the IFU data cube and take the square root of the result. This gives us the noise coefficients in wavelet space.\\
\\
In the next step, we build the multiresolution support \textbf{\textsf{M}}, which is a $(L+1) \times P$ matrix. We compare the absolute value of the signal and noise wavelet coefficients at each pixel, $w_{s,l}$ and $w_{s,l}^N$. We take a threshold $T$ set by the user, for example 5 for a 5$\sigma$ cleaning in wavelet space, and set the corresponding matrix entry in \textbf{\textsf{M}} to 1 if $|w_{s,l}| \geq T |w_{s,l}^N|$, and 0 otherwise. Note that for $s=1$, we use a higher threshold of $T+1$, as this wavelet scale corresponds to high frequencies, where we expect the noise to dominate. The matrix coefficients for the smoothed signal $\boldsymbol{c_L}$ are automatically set to 1.\\
\\
Now we perform the cleaning: We set all $w_{s,l}$ associated with a vanishing \textbf{\textsf{M}} value to zero and transform back into real space to obtain a first clean spectrum. However, there is still some signal to be harnessed in the residuals. Therefore we subtract the clean spectrum from the full spectrum to obtain the residual spectrum, and we compare its standard deviation, $\sigma_{\text{res}}$, with the standard deviation of the full spectrum (in the first iteration) or of the residual used in the previous iteration (all subsequent iterations), which we indicate in both cases with $\sigma_{\text{prev}}$. If $|(\sigma_{\text{prev}} - \sigma_{\text{res}})/\sigma_{\text{res}}| > \epsilon$, we transform the residual spectrum into wavelet space, set wavelets with vanishing \textbf{\textsf{M}} values to zero, transform back into real space, and add the resulting signal to obtain our new clean signal. Note that the same multiresolution support as before is used. Subsequently, we calculate again the residual and continue until the $\epsilon$ criterion is no longer fulfilled and all the signal has been extracted. The value of $\epsilon$ is set by the user and must satisfy the condition $0 < \epsilon < 1$. Algorithm~\ref{algorithm:Cleaning} summarizes this cleaning procedure.
\\
\begin{algorithm}
	\renewcommand{\hypcapspace}{30pt} 
	\capstart 
	\renewcommand{\hypcapspace}{0.5\baselineskip} 
	\begin{algorithmic}[1]
		\Statex \textbf{Require:} Wavelet transform of spectrum $\{\boldsymbol{w_1}, \dots, \boldsymbol{w_L}, \boldsymbol{c_L}\}$, highest scale level $L$, 
		number of spectrum pixels $P$,  convolution mask $\boldsymbol{h}$
		\Statex \textbf{Output:} Spectrum in real space $\boldsymbol{c_0}$ (= set of discrete spectrum pixels $\{c_{0,l}\}$)
		\State \textbf{Procedure:} \textsc{wavelet\_backtransform}($\{\boldsymbol{w_1}, \dots, \boldsymbol{c_L}\}$)\textbf{:}
		\State $\boldsymbol{S} \gets \boldsymbol{c_L}$
		\ForAll{$s \in \{1, \dots, L\}$}
		\ForAll{$l \in \{1, \dots, P\}$}
		\State $C_l \gets \sum_k h(k)~S_{l+2^{L-s}k}$
		\EndFor
		\State $\boldsymbol{S} \gets \boldsymbol{C} + \boldsymbol{w_{L+1-s}}$
		\EndFor
		\State $\boldsymbol{c_0} \gets \boldsymbol{S}$ 
		\State \Return $\boldsymbol{c_0}$
	\end{algorithmic}
	\caption{Transformation from wavelet to real space}\label{algorithm:WaveletToSignalSpace}
\end{algorithm} 
\\
\\
\begin{algorithm}
\renewcommand{\hypcapspace}{30pt} 
\capstart 
\renewcommand{\hypcapspace}{0.5\baselineskip} 
\begin{algorithmic}[1]
\Statex \textbf{Require:} Spectrum $\boldsymbol{c_0}$ (= set of discrete spectrum pixels $\{c_{0,l}\}$), $\boldsymbol{\sigma}_{\textbf{spec}}$ ( = vector with standard deviation noise estimate for each spectrum pixel), 
highest scale level $L$, number of spectrum pixels $P$,  cleaning threshold $T$, cleaning parameter $\epsilon$ ($0 < \epsilon < 1 $)
\Statex \textbf{Output:} Cleaned spectrum $\boldsymbol{S}_{\textbf{clean}}$
\State \textbf{Procedure} \textsc{clean\_signal}($\boldsymbol{c_0}$, $\boldsymbol{\sigma}_{\textbf{spec}}$, $L$, $T$, $\epsilon$)\textbf{:}
\State $\{\boldsymbol{w_{1}^{\delta}}, \dots, \boldsymbol{w_{L}^{\delta}},\boldsymbol{ c_{L}^{\delta}}\} \gets$ \textsc{wavelet\_transform}($\delta$-dist., $L$)
\ForAll{$\boldsymbol{w_s^{\delta}} \in \{\boldsymbol{w_{1}^{\delta}}, \dots, \boldsymbol{w_{L}^{\delta}}\}$}
\State $\boldsymbol{w_s^N} \gets \sqrt{\boldsymbol{w_s^{\delta}}{}^{2} \ast \boldsymbol{\sigma}_{\textbf{spec}}^2 }$
\EndFor
\State $\{\boldsymbol{w_1}, \dots, \boldsymbol{w_L}, \boldsymbol{c_L}\} \gets$ \textsc{wavelet\_transform}($\boldsymbol{c_0}$, $L$)
\State $\textbf{\textsf{M}} \gets \textbf{\textsf{0}}_{L+1,P}$ \hfill // Multiresolution support matrix
\ForAll{$s \in \{1, \dots, L+1\}$, $l \in \{1, \dots, P\}$}
\If{$s == 1~ \textbf{and}~ |w_{s,l}| \geq (T+1) |w_{s,l}^N|$}
\State $M_{sl} \gets 1$
\ElsIf{$1 < s \leq L~ \textbf{and}~ |w_{s,l}| \geq T |w_{s,l}^N|$}
\State $M_{sl} \gets 1$
\ElsIf{$s == L+1$}
\State $M_{sl} \gets 1$
\EndIf
\EndFor
\State $\boldsymbol{S}_{\textbf{clean}} \gets \boldsymbol{0}_P$, $\sigma_{\text{prev}} \gets 0$, $\textbf{\textit{res}} \gets \boldsymbol{c_0}$
\State $\sigma_{\text{res}} \gets \text{std}(\textbf{\textit{res}})$
\While{$|(\sigma_{\text{prev}} - \sigma_{\text{res}}) / \sigma_{\text{res}}| > \epsilon$}
\State $\{\boldsymbol{w_{1}^{\textbf{res}}}, \dots, \boldsymbol{c_L^{\textbf{res}}}\} \gets$ \textsc{wavelet\_transform}(\textbf{\textit{res}}, $L$)
\ForAll{$s \in \{1, \dots, L\}$, $l \in \{1, \dots, P\}$}
\If{$M_{sl} == 0$}
\State $w_{s,l}^{\text{res}} \gets 0$
\EndIf
\EndFor
\State $\textbf{\textit{res}}_{\textbf{clean}} \gets$ \textsc{wavelet\_backtransform}($\{\boldsymbol{w_{1}^{\textbf{res}}}, \dots, \boldsymbol{c_L^{\textbf{res}}}\}$)
\State $\boldsymbol{S}_{\textbf{clean}} \gets \boldsymbol{S}_{\textbf{clean}} + \textbf{\textit{res}}_{\textbf{clean}}$
\State $\textbf{\textit{res}} \gets \boldsymbol{c_0} - \boldsymbol{S}_{\textbf{clean}}$
\State $\sigma_{\text{prev}} \gets  \sigma_{\text{res}}$
\State  $\sigma_{\text{res}} \gets \text{std}(\textbf{\textit{res}})$
\EndWhile
\State \Return $\boldsymbol{S}_{\textbf{clean}}$
\end{algorithmic}
\caption{Signal cleaning in wavelet space}\label{algorithm:Cleaning}
\end{algorithm}

\subsection{Testing the wavelet-based reconstruction}
\begin{figure}
\begin{center}
\includegraphics[scale=0.18]{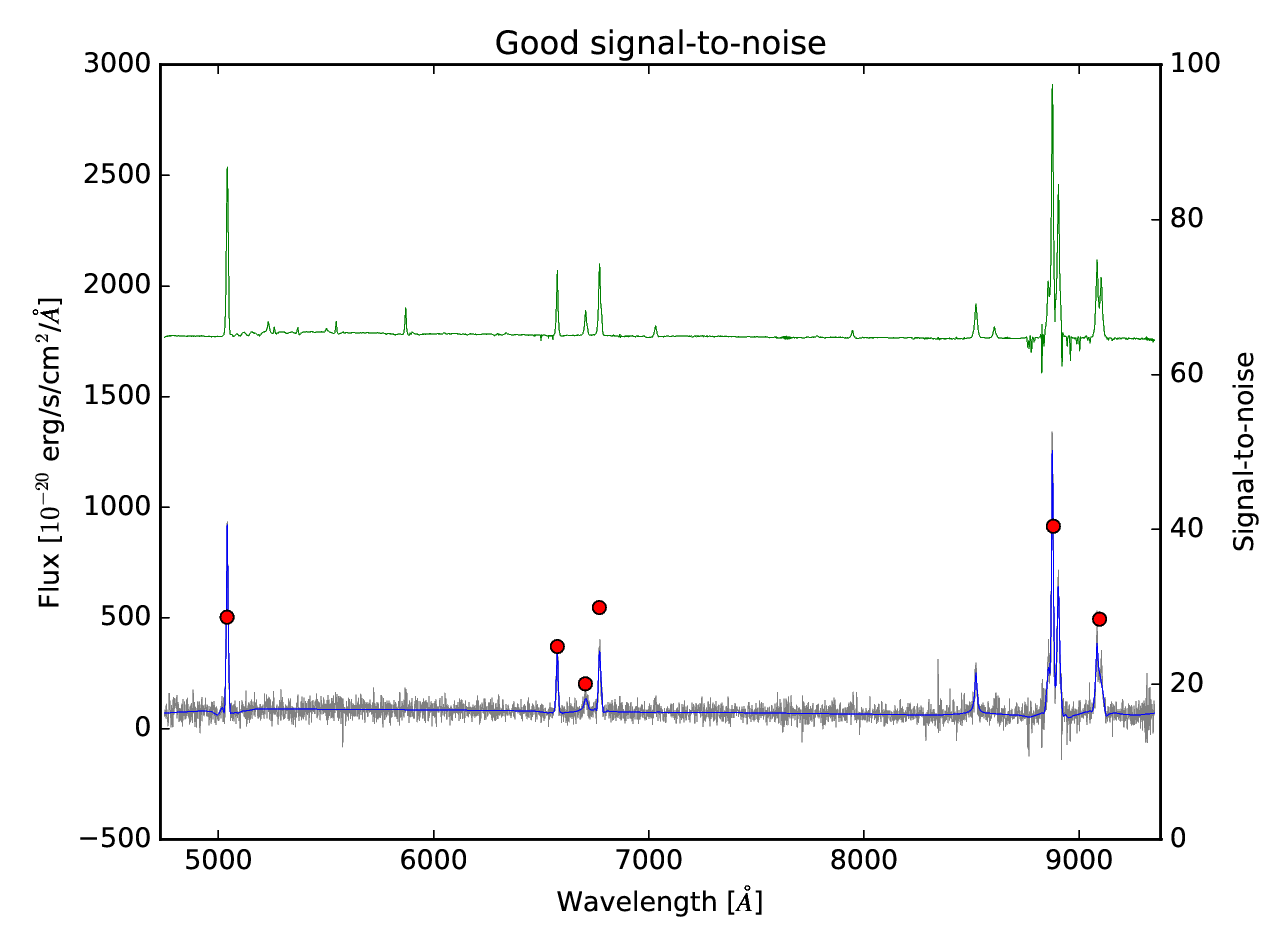}
\includegraphics[scale=0.18]{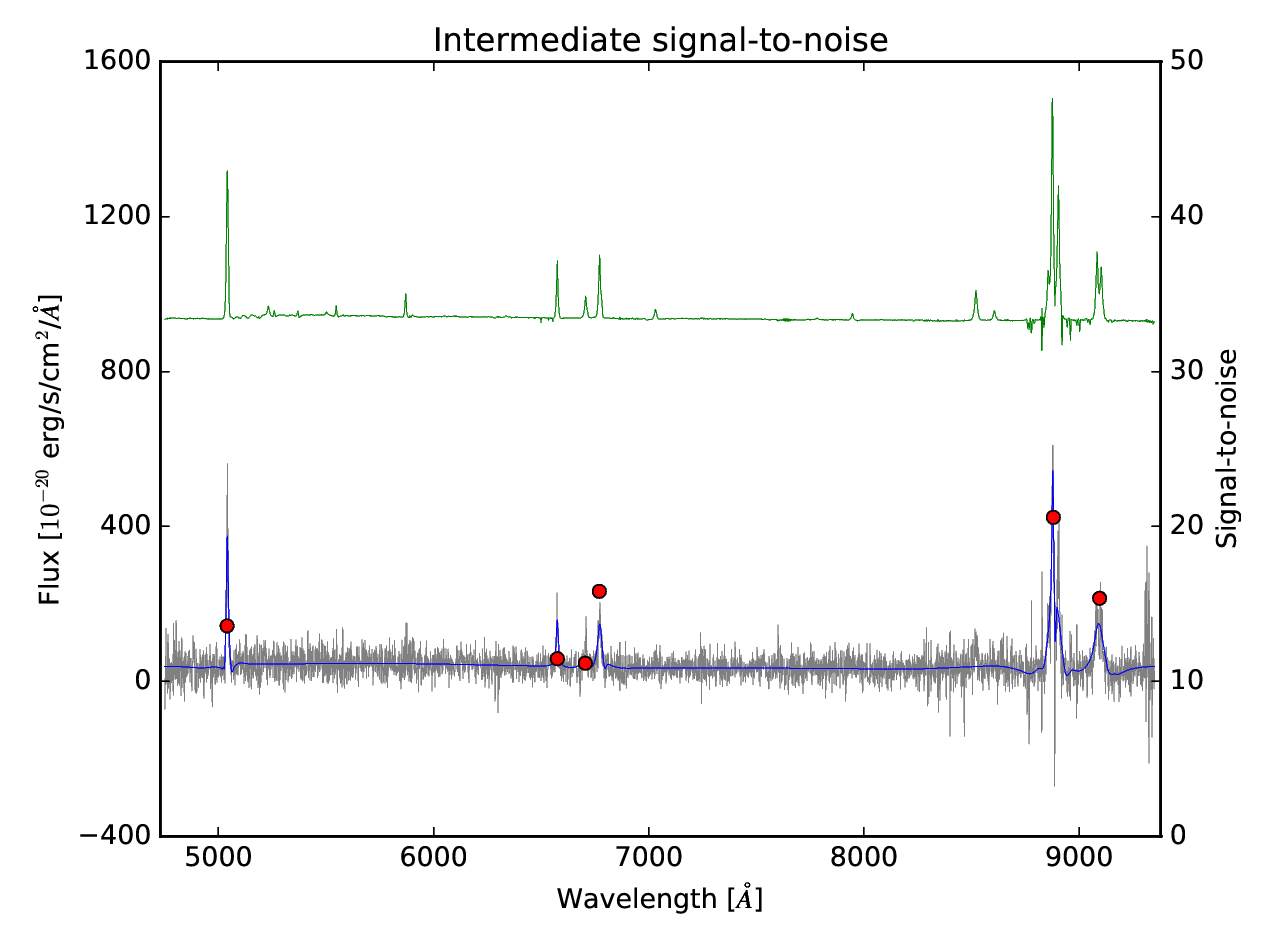}
\includegraphics[scale=0.18]{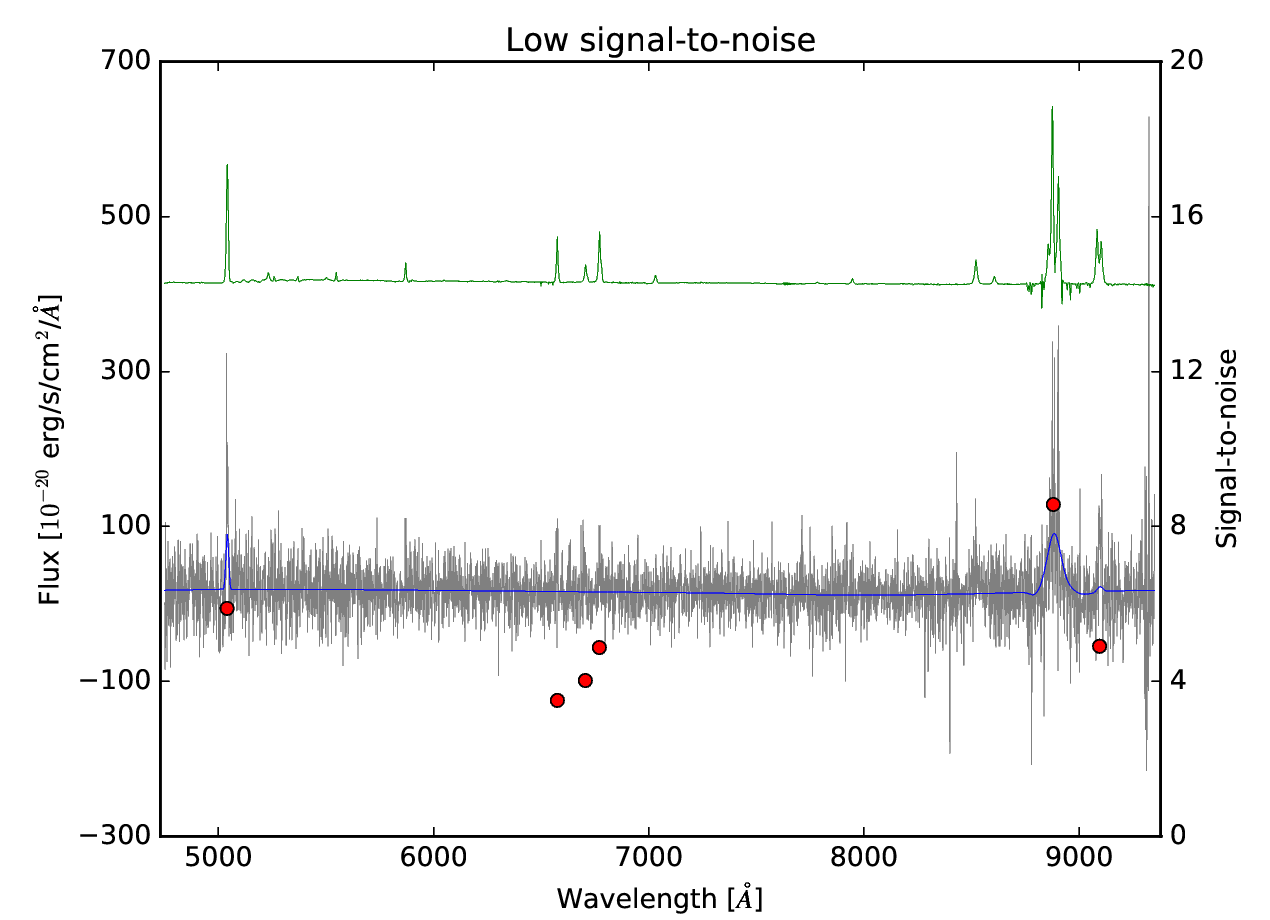}
\caption{Reconstructed spectra for three different signal-to-noise scenarios using 5$\sigma$ cleaning and $\epsilon = 0.01$. The true spectrum is offset and shown in green, the noisy spectrum is displayed in gray and the reconstructed signal in blue. Red dots indicate the signal-to-noise of the respective emission lines. The wavelet tool detects peaks with $S/N \gtrsim 8$. However, if a low $S/N$ emission line is located very close to a high $S/N$ line, it is possible that it will not be recovered (middle plot, 3rd emission line from the left). The wavelet reconstruction can occasionally even find lines with $S/N < 8$ (bottom plot, 1st and 6th emission line), but typically they will not be detected (bottom plot, 2nd - 4th emission line).}\label{figure_waveletCleaningTest}
\end{center}
\end{figure}

\begin{figure}
\begin{center}
\includegraphics[scale=0.18]{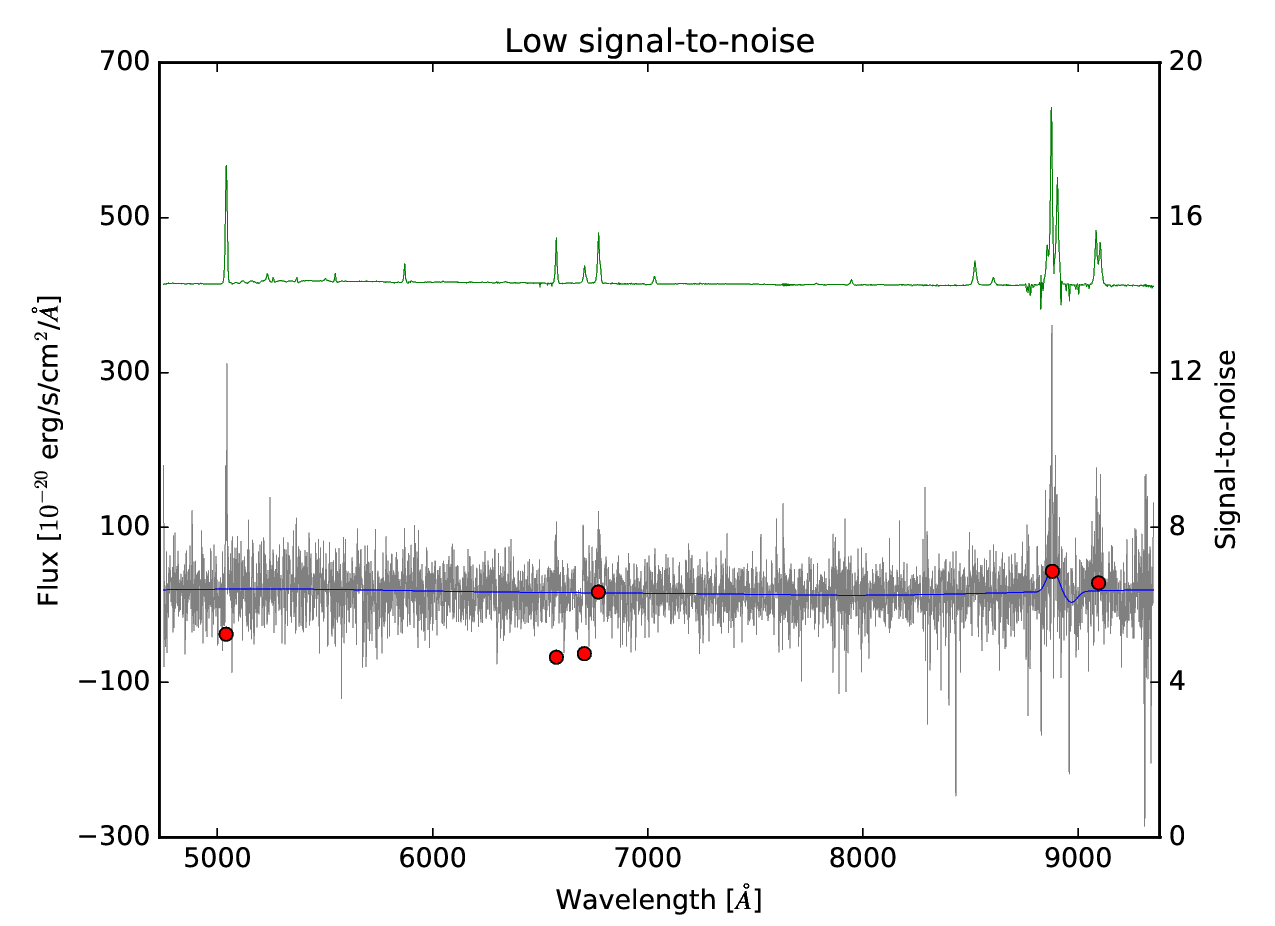}
\caption{In low signal-to-noise (S/N $\leq 10$) cases, ringing effects can occasionally lead to signal dips with similar amplitude as the signal peaks of the reconstructed emission line. Therefore care has to be taken not to mistake these effects for absorption lines. This might be ameliorated by re-running the wavelet reconstruction with a different setup. The colors have the same meaning as in figure~\ref{figure_waveletCleaningTest} and the reconstruction was performed with a 5$\sigma$ cutoff and $\epsilon=0.01$.}\label{figure_ringingEffect}
\end{center}
\end{figure}

To test our software, we use the spectrum of the brightest cluster galaxy (BCG) from our MUSE data set described in the next section. MUSE provides both the spectrum signal and a noise estimate over the full wavelength range. The original spectrum can be considered clean due to its very high signal-to-noise. We rescale it to simulate fainter sources at low signal-to-noise. We calculate the rescaling factor $R$ by looking at the highest spectrum signal peak and dividing the associated MUSE noise estimate by this signal. This results in $R \approx 0.0015$. We investigate three cases, namely a good, an intermediate, and a low signal-to-noise case, where we rescale the full signal spectrum by $10 R$, $5 R$, and $2 R$ respectively. Subsequently we add Gaussian noise simulating the real noise estimate of the MUSE data cube. For each spectral wavelength pixel $l$ we obtain the realized noise by drawing from a Gaussian probability distribution with a standard deviation equal to the MUSE standard deviation noise estimate at this pixel. We repeat this process 10 times to obtain spectra with different noise realizations.\\
\\
We calculate the signal-to-noise of six emission lines by summing over their respective wavelength ranges,
\begin{equation}
\frac{\text{S}}{\text{N}} = \frac{\sum_l \text{signal}_l}{\sqrt{\sum_{l'}\text{std}^2_{l'}}},
\end{equation}
where $\text{std}_l$ is the MUSE standard deviation noise estimate at pixel $l$. We will refer to the lines according to their wavelength order, i.e. the first line is situated at the lowest wavelength and the last line at the highest. We apply our wavelet cleaning software to the spectra using the MUSE noise estimate and different wavelet parameters as input. We investigate 5$\sigma$ and 3$\sigma$ cleaning and $\epsilon$ parameters of 0.1, 0.01, and 0.001. The cleaning procedure is fast and takes about 1 second per spectrum on a laptop. Figure~\ref{figure_waveletCleaningTest} shows reconstructed spectra for the three different signal-to-noise cases. Note that the last two emission lines in the true spectrum are actually comprised of merged individual lines. As can be seen in figure~\ref{figure_waveletCleaningTest}, the wavelet tool can detect if a line consists of two merged lines and reconstruct them correctly if their signal-to-noise is high enough. If it is too low, it will reconstruct them as a single line.   
\\
\\
For all 90 spectra which we analyzed with a 5$\sigma$ wavelet reconstruction, we find no fake detections of emission lines. For signal-to-noise larger than 20, all 6 test emission lines are detected. For S/N between 10 and 20, all emission lines but the third are found. The third peak is no longer recovered due to its proximity to the fourth peak, which has typically a twice larger S/N value. In general, the wavelet software might reconstruct two close-by peaks as a single peak unless they have each a sufficiently large signal. When the signal-to-noise of both peaks was similar, both the third and the fourth emission line were detected and reconstructed. For emission lines with low signal-to-noise values between 5 and 10, we can reconstruct the stronger lines with S/N $\gtrsim 8$, while the weaker peaks remain typically undetected. However, as the bottom plot in figure~\ref{figure_waveletCleaningTest} shows, even weaker peaks can occasionally be reconstructed.\\
\\
Emission lines modeled with a wavelet reconstruction do sometimes not reach the full peak height of the signal, in particular for high $\epsilon$ values, and their tails can suffer from ringing effects which might be due to the wavelet shape, see for example the first emission line of the intermediate S/N case in figure~\ref{figure_waveletCleaningTest}. For low signal-to-noise emission lines (S/N $\leq 10$), care has therefore to be taken not to mistake the signal dip due to ringing effects as an absorption signal, as the ringing effect might occasionally have a similar (negative) amplitude as the signal peak of the reconstructed emission line, see figure~\ref{figure_ringingEffect}. When this effect occurs in practice, it might be improved by changing the wavelet setup, e.g. by lowering the $\epsilon$ value. A lower $\epsilon$ is designed to detect a larger fraction of the signal peak and should thus increase its height. However, care has to be taken as a lower $\epsilon$ might also lead to stronger ringing effects.\\
\\
\\
The 3$\sigma$ wavelet reconstruction recovered more emission lines than the 5$\sigma$ cleaning, but it also produced false detections. We therefore adopted a conservative approach and used the 5$\sigma$ wavelet cleaning when applying the code to real data.\\
\\
Finally, we compared the noise free emission line shapes with the reconstructed ones. We find that the shape reconstruction is generally good, but the reconstructed line shape and height recovered from the noisy data can differ from the original, clean ones, in particular in low signal-to-noise scenarios. Therefore we use the wavelet cleaning only to distinguish true from false spectrum peaks, and we perform all data operations such as fitting a Gaussian to obtain the centering error on the real, noisy data.

\section{Application to MUSE data: MACSJ1931}
\begin{figure}
\begin{center}
\includegraphics[width=0.5\textwidth]{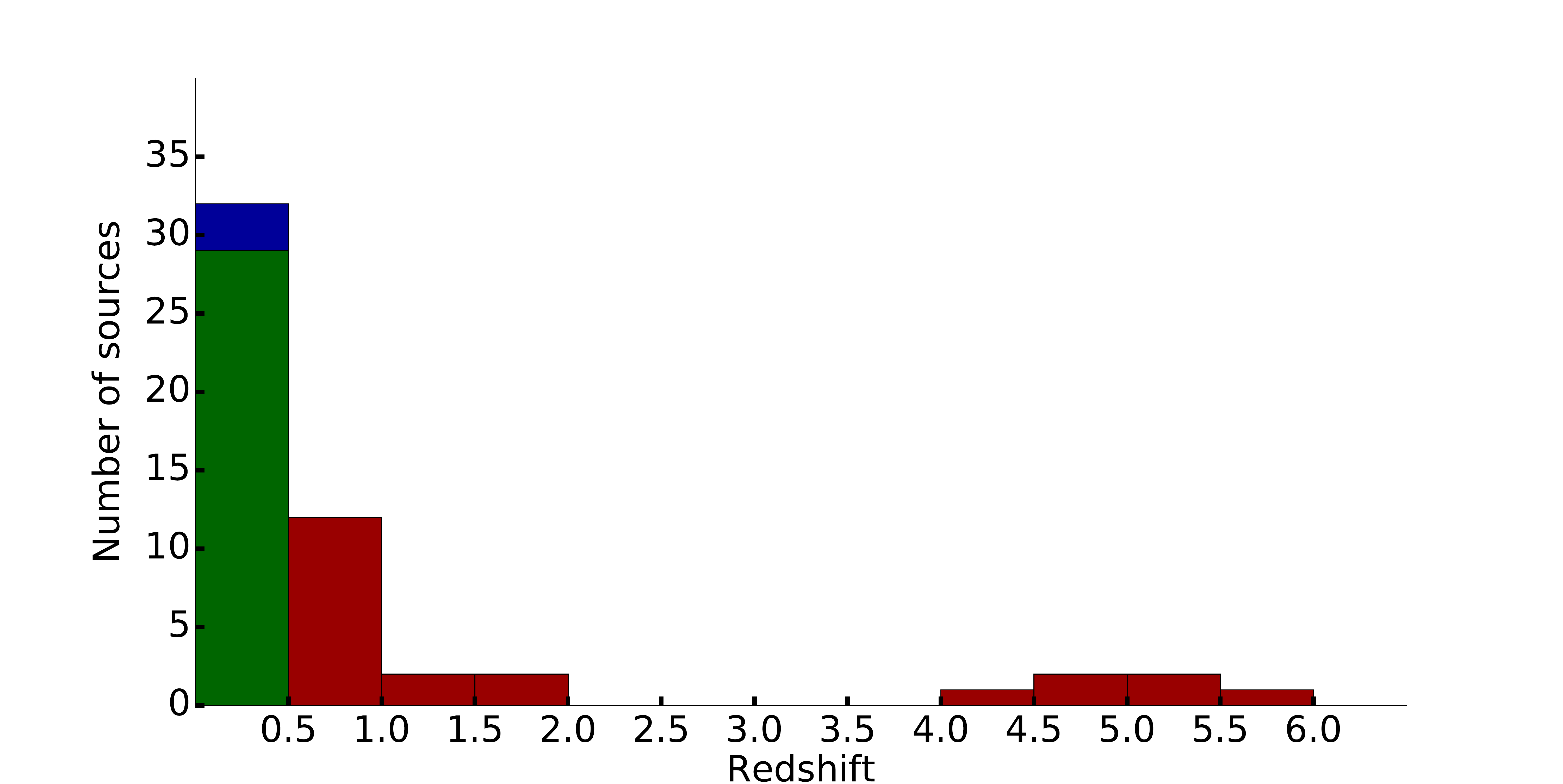}
\caption{Source distribution in redshift space. Background sources are colored in red, cluster members in green and the remaining objects in blue.}\label{fig:redshift_histogram}
\end{center}
\end{figure}

We apply IFS-RedEx to our data set of the strong lensing cluster MACSJ1931 obtained with MUSE \citep{Bacon2010a} on the Very Large Telescope (VLT). We combine our data with the publicly available \textit{Hubble Space Telescope} (HST) imaging from the Cluster Lensing And Supernova survey with Hubble \citep[CLASH, ][]{Postman2012}. The cluster is part of the MAssive Cluster Survey (MACS), which comprises more than one hundred highly X-ray luminous clusters \citep{Ebeling2010,Ebeling2001}.\\
\\
The core of MACSJ1931 ($z = 0.35$) was observed with MUSE on June 12 and July 17 2015 (ESO program 095.A-0525(A), PI: Jean-Paul Kneib). The 1 x 1 arcmin$^2$ field of view was pointed at $\alpha$ = 19:31:49.66 and $\delta$ = -26:34:34.0 (J2000) and we observed for a total exposure time of 2.44 hours, divided into 6 exposures of 1462 seconds each. We rotated the second exposure of each exposure pair by 90 degrees to allow for cosmic ray rejection and improve the overall image quality. The data were taken using the WFM-NOAO-N mode of MUSE in good seeing conditions with FWHM $\approx 0.7$ arcseconds.\\
\\
We reduced the data using the MUSE pipeline version 1.2.1 \citep{Weilbacher2014,Weilbacher2012}, which includes bias and flat-field corrections, sky subtraction, and wavelength and flux calibrations. The six individual exposures were finally combined into a single data cube and we subtracted the remaining sky residuals with ZAP \citep{Soto2016}. The wavelength range of the data cube stretches from 4750 to 9351~\AA~in steps of 1.25~\AA. The spatial pixel size is 0.2 arcseconds.\\
\\
We used the HST data for MACSJ1931 obtained as part of the CLASH program \citep{Zitrin2015} in the bands F105W, F475W, F625W, and F814W with a spatial sampling of 0.03 arcsec/pixel. The HST data products are publicly available on the CLASH website\footnote{\url{https://archive.stsci.edu/prepds/clash/}}. \\
\\
We use only redshift identifications which we consider secure because we see e.g. several lines or a clear Ly$\alpha$ emission line shape. We extract 54 sources with redshifts ranging from 0.21 to 5.8. Among them, 29 are cluster members with $0.3419 \leq z \leq 0.3672$ and 22 are background sources with $0.4 \leq z \leq 5.8$. A table of all sources with spectroscopic redshifts is presented in the companion paper Rexroth et al. 2017 (in preparation), in which we use the data to improve the cluster lens model. Figure~\ref{fig:redshift_histogram} shows a histogram of the source distribution in redshift space.

\section{Summary}
We describe IFS-RedEx, a public spectrum and redshift extraction pipeline for integral-field spectrographs. The software supports SExtractor catalogs as well as MUSELET narrow-band detection catalogs as input. The pipeline has several features which allow a quick identification of reliable spectrum features, most notably a wavelet-based spectrum cleaning tool. The tool only reconstructs spectral features above a given significance threshold. We test it with degraded MUSE spectra and find that it can detect spectral features with $\text{S/N} \gtrsim 8$. We find no fake detections in our test. Finally, we apply IFS-RedEx to a MUSE data cube of the strong lensing cluster MACSJ1931 and extract 54 spectroscopic redshifts.

\section*{Acknowledgements}
MR thanks Timoth\'{e}e Delubac for verifying the spectral line identifications and Yves Revaz and the ESO user support center for their help with a non-critical issue in the MUSE pipeline. He thanks Thibault Kuntzer, Pierre North and Fr\'{e}d\'{e}ric Vogt for fruitful discussions and Anton Koekemoer for his help with processing the Simple Imaging Polynomial (SIP) distortion information from FITS image headers. MR and JPK gratefully acknowledge support from the ERC advanced grant LIDA. RJ gratefully acknowledges support from the Swiss National Science Foundation. JR gratefully acknowledges support from the ERC starting grant 336736-CALENDS. This research made use of SAOImage DS9, numpy \citep{Walt2011}, scipy \citep{SciPyModule}, matplotlib \citep{Hunter2007}, PyFITS, PyRAF/IRAF \citep{Tody1986}, Astropy \citep{AstropyCollaboration2013}, pyds9, GPL ghostscript, and TeX Live. PyRAF and PyFITS are public software created by the Space Telescope Science Institute, which is operated by AURA for NASA. This research has made use of NASA's Astrophysics Data System.

\bibliography{Astronomy_papers.bib}

\bsp
\label{lastpage}
\end{document}